\documentclass[12pt,a4paper]{article}
\usepackage{amssymb}
\usepackage{bezier}
\usepackage{emlines2}
\usepackage[T2A]{fontenc}
\usepackage[cp866]{inputenc}
\usepackage[russian,english]{babel}
\normalbaselines
\begin{document}
\voffset=-2cm
\begin{center}
{\LARGE \bf {Is classical reality completely deterministic?}}\\[5mm]
{\bf B.\ P.\ Kosyakov}\footnote{{\it E-mail address}: 
${\rm kosyakov@vniief.ru}$}\\[3mm]
{Russian Federal Nuclear Center, Sarov, 607190 Nizhnii
Novgorod Region, Russia} 
\end{center}
\begin{abstract}
\noindent
The concept of determinism for a classical system is interpreted 
as the 
requirement that the solution to the Cauchy problem for the equations of 
motion governing this system be unique.
This requirement is generally assumed to hold for all autonomous 
classical systems. 
We give 
counterexamples of this view.
Our analysis of classical electrodynamics in a 
world with one temporal and one spatial dimension shows that the solution to 
the Cauchy problem with the initial conditions of a particular
type is not unique.
Therefore, random behavior of closed classical systems is indeed possible.
This finding provides a qualitative explanation of how classical strings can split. 
We propose a modified path integral formulation of classical mechanics  
to include indeterministic systems.
\end{abstract}
\vskip2mm
\section{Introduction}
According to present views, the quantum is fundamentally random.
By this is meant that quantum mechanics is a probabilistic theory 
and there are no 
deterministic laws underlying quantum phenomena.
By contrast, the classical is regarded as deterministic.
Of course,  
classical statistical 
mechanics invokes probability theory,
but the reason for this is different from that of quantum mechanics.
Uncertainties in classical statistical mechanics may be attributed 
to lack of knowledge of
actual deterministic histories of macroscopic systems which have too many 
degrees of freedom to be completely controlled. 
 
Worthy of mention are also classical stochastic systems
(among which are systems with  
some few degrees of freedom) \cite{Tabor}.
Although stochastic mechanics is formulated with the help of probability theory, 
stochasticity should not be confused with randomness.
Classical stochastic systems are governed by deterministic laws. 
The gist of the question is that their histories are depicted by 
{tangled} trajectories.
Motions displaying extreme sensitivity to initial conditions are
commonly {viewed} as stochastic.
Complexity effects in the behavior of 
unstable systems are a major manifestation of stochasticity.
To be more exact, a system is defined as stochastic if there is a compact region 
confining the motion $x(t)$ in which $x(t)$ depends heavily on initial data  
$x_0$:
\begin{equation}
\frac{\partial x(t;x_0)}{\partial x_0}\sim \exp\left({t}/{\Delta}\right), 
\quad
t\gg{\Delta} 
\label
{stochastic}
\end{equation}                                            
(where ${\Delta}$ stands for a characteristic time interval).
The apparent indeterminism in the behavior of  stochastic systems is then 
fictitious; it is due to 
imperfect knowledge of initial conditions.
We can in principle specify $x_0$ with arbitrary accuracy,
and thereby predict the history $x(t;x_0)$ as precisely as desired.

To discern phenomena which indeed run counter to Laplace's determinism, we
must refine upon this paradigm.
We say that Laplace's determinism holds for a given 
system if the Cauchy problem for the equations of motion governing this 
system---whenever the initial conditions---has a {\it unique} solution.
This requirement is generally believed to be imperative in classical 
physics. 
Strange as it may seem, there are autonomous classical systems in 
two-dimensional spacetime which violate 
this principle.
Examples of such indeterministic systems  are given below.
We will see that behavior of these systems must be recognized as truly random.
In two-dimensional worlds, God does roll the dice.
 
It may be that this implication will have some utility in string theory.
By now, there has been remained an open question of whether 
fundamental strings 
can split on the classical level.
At first sight, classical strings are unable to split at all.
Take, for example, an open Nambu string.
It can be indefinitely stretched without no evidence of being favorably
disposed towards splitting\footnote{See, however, Ref. \cite{deVega} where
a thermodynamical argument in support of the idea that some classical 
string configurations show tendency to split is adduced.
It seems appropriate to reason that some splitted configuration is more 
advantageous in mass content than 
its associated unbroken configuration, but this criterion is in general insufficient 
to determine the point of the string where splittng actually occurs.}:
there is no elastic limit
for objects  governed by the Nambu action.
Indeed, the only dimensional parameter in this action is $1/2\pi{\alpha}'$
which is merely an overall 
factor that defines 
the scale of length. 
It follows that classical strings are immune from compulsory splittings.
However, as will transpire in Sect. 3, {\it spontaneous}
splittings are yet 
feasible in the classical picture.

The paper is organized as follows.
In Sect. 2 we explore a particle on the top of a hill.
From this discussion, a general idea can be had of how 
classical systems can reveal its indeterministic nature.
Classical electrodynamics of point particles in a 
world with one temporal and one spatial dimension is analyzed in Sect. 3.
We show that exact solutions to the Cauchy problem for the set of 
dynamical equations governing a closed system of two charged particles 
and the electromagnetic field can be not unique.
We then propose a toy model which qualitatively explains random splitting 
of classical strings. 
In the final section, we turn to 
the path integral formulation of classical mechanics 
\cite{gozzi}--\cite{persik}.
If indeterministic systems are to be incorporated,
the classical path integral construction should be properly modified. 
We outline a possible way for this modification.

\section{At the top of a hill}
Let us take a closer look at two like charged particles which move 
towards each other along a straight line.
Having spent kinetic energy for overcoming the interparticle
repulsion by their meeting these particles 
merge into a single point aggregate.
Since our concern is with final stage of this head-on collision 
when velocities of 
the particles are close to zero, the use of 
nonrelativistic approximation  
would be quite accurate.

\begin{figure}[htb]
\begin{center}
\unitlength=1.00mm
\special{em:linewidth 0.4pt}
\linethickness{0.4pt}
\begin{picture}(90.00,30.00)
\bezier{180}(10.00,10.00)(20.00,30.00)(30.00,10.00)
\emline{40.00}{10.00}{1}{50.00}{20.00}{2}
\emline{60.00}{10.00}{3}{50.00}{20.00}{4}
\bezier{64}(70.00,10.00)(77.00,14.00)(80.00,20.00)
\bezier{64}(90.00,10.00)(83.00,14.00)(80.00,20.00)
\put(10.00,13.00){\vector(2,3){3.00}}
\put(10.00,13.00){\makebox(0,0)[cc]{$\bullet$}}
\put(41.00,13.00){\vector(1,1){4.00}}
\put(41.00,13.00){\makebox(0,0)[cc]{$\bullet$}}
\put(70.00,12.00){\vector(4,3){4.00}}
\put(70.00,12.00){\makebox(0,0)[cc]{$\bullet$}}
\put(20.00,5.00){\makebox(0,0)[cc]{$a$}}
\put(50.00,5.00){\makebox(0,0)[cc]{$b$}}
\put(80.00,5.00){\makebox(0,0)[cc]{$c$}}
\end{picture}
\caption{Ascent to the top of a hill}
\label
{hill-g}
\end{center}
\end{figure}
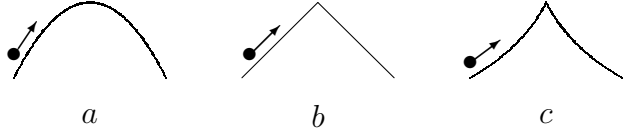

The two-particle problem can be brought to a one-particle problem 
if we introduce the relative coordinate $r=x_2-x_1$, reduced mass 
$m=m_1 m_2/(m_1+m_2)$, and potential energy 
$U(r)$.
The problem is then to describe a particle climbing to the hill 
$U(r)$ so that its velocity vanishes on its arrival at the top of the hill, 
see Fig. \ref{hill-g}.
Let the coordinate of the top be $r=0$, $U_{\rm max}=U(0)$.
The time that the particle comes to the top is chosen 
to be  $t=0$.
Vanishing the particle's velocity at $r=0$ means that the total 
energy is zero, 
\begin{equation}
E={\frac12}\,{m}{\dot r}^2+U(r)=0.
\label
{hill}
\end{equation}                                            
The time it takes for the particle to arrive at the top is therefore
\begin{equation}
t(r)=-\sqrt{\frac{m}{2}}\int_r^0 \frac{dx}{\sqrt{-U(x)}}\,.
\label
{t}
\end{equation}                                            

What happens to the particle upon its arrival at the top?
(Although this question seems sound, no attempt has been made of 
answering it.
The only exception is Ref. \cite{Macomber}, where
this issue is addressed, but its solution is left to 
the reader\footnote{Subsequent to posting the first version of this 
paper, I was informed by Prof. J. Norton that similar problem
was discussed in his paper \cite{norton} with an emphasis on 
philosophical aspects.}.)
If the integral in (\ref{t}) diverges, then the question of 
the subsequent lot of this particle does not arise because the ascent takes 
infinite time. 
Such is indeed the case for $U(r)$ which is an analytical function at $r=0$,
say, for $U(r)=-U_0\,r^2$.
Meanwhile the integral is finite for 
\begin{equation}
U(r)= -U_0\, r^{2(1-\alpha)},\quad 0<\alpha<1,
\label
{deg}
\end{equation}                                            
\begin{equation}
U(x)= -U_0\,  r^2\left(\ln r^2\right)^{2(1+\beta)},\quad \beta>0,
\label
{ln}
\end{equation}                                            
\begin{equation}
U(r)= -U_0\,  r^2\left(\ln r^2\right)^2\left[\ln\left(\ln r^2\right)^2\right]^{2(1+\gamma)},
\quad\gamma>0,
\label
{lnln}
\end{equation}                                            
{and the like}.

The differential equation (\ref{hill}) is invariant under time 
reversal.
Furthermore, $r(t)=0$ is another solution to (\ref{hill}).
Therefore, if the climb takes a finite period of time, then
a continual set of options is available: 
after staying at the top for an 
arbitrary period of time $T$, the particle can start to descend in either 
direction\footnote{If we bear in mind the initial two-particle problem and 
assume that colliding particles are unable to penetrate through each other, 
then both ascent and descent in the effective one-particle problem 
are described by positive values of $r$.}.
Analytically, 
\begin{equation}
r(t)=\cases{f(t) & $t<0$,\cr
0 & $0\le t< T$,\cr
\pm\,f(T-t) & $t\ge T$,\cr}
\label
{e-}
\end{equation}                                           
where $f(t)$ is the inverse of $t(r)$ defined in (\ref{t}). 

By the Picard theorem, the solution to the Cauchy problem for the differential
equation (\ref{hill}) 
with the initial condition $r=0$ 
is unique if the Lipschitz condition holds,
\begin{equation}
\sqrt{-U(r)}< C\,\vert\,r\,\vert.
\label
{lip}
\end{equation}                                            
Here, $C$ is some positive constant.
Clearly, for $U(r)$ given by  
(\ref{deg})--(\ref{lnln}), 
inequality (\ref{lip})  fails. 

We thus infer that potentials, which are visualized as hills, 
are divided into two classes:
{unstable potentials of the conventional type} and 
{\it over-unstable} potentials.
The equilibrium state in potentials of the conventional type is 
kept until a small external 
perturbation occurs, whereas 
this state in over-unstable potentials can be upset {\it spontaneously}, 
that is, with no external cause.

The Lipschitz condition is sufficient but not necessary for
stability against spontaneous decays.
Convergence of the integral in (\ref{t}) may serve a necessary 
condition.
For example, inequality (\ref{lip}) does not hold for 
$U= -U_0\,r^2\left(\ln r^2\right)^2$, 
even if the integral in (\ref{t}) diverges.
Note also the absence of a strict analytical demarcation line between 
unstable potentials of the conventional type  and over-unstable potentials,
in particular, the sequence of over-unstable potentials
shown in (\ref{deg})--(\ref{lnln})
extends indefinitely.

A striking thing is that a particle at rest shows the capacity 
for sliding down the hill {without any causation} and thus {at random}.
This phenomenon is in conflict with Laplace's determinism.
Going back to the initial two-particle problem, we see that
the aggregate of two merged particles will spontaneously disintegrate
 into its 
constituents after a lapse of an arbitrary interval $T$.
Note that $T=\infty$ is among possible options, that is to say
the aggregate can remain fixed for an  infinitely long time.

The sceptical reader may disregard these issues for several compelling reasons.
First,  head-on collisions of point particles are highly improbable
on a three-dimensional arena:
the probability measure of such events is zero.
Second, the interaction potential $U(r)$ like that shown in 
(\ref{deg})--(\ref{lnln}) seems to have
little (if any) significance as an element of physical reality.
Third, time reversal is crucial for spontaneous equilibrium breaking
to occur.
Once accelerated charges radiate electromagnetic energy, the dynamics 
becomes dissipative and irreversible, and
hence solution (\ref{e-}) ceases to be true.
Fourth, to ensure that two colliding particles amalgamate in a single aggregate,
their total energy must be exactly zero.
The initial data of the corresponding Cauchy problems constitute a null set.

All these objections can be withdrawn if we turn to
a world with one temporal and one spatial dimension.
First, observe that, for particles living in a line, head-on collisions are 
not uncommon.
Second, with reference to \cite{k9, k2006}, we recall that the time component of the 
retarded vector potential $A_\mu$ in two-dimensional electrodynamics  
is given by $A_0=-e\,\vert\,r\,\vert$  which, 
on putting $\alpha=\frac12$, falls into the type of 
(\ref{deg}), Fig. \ref{hill-g}$b$. 
Third, it was shown in \cite{k9, k2006} that charged particles in two-dimensional
spacetime do not radiate.
Therefore, all processes in this realm are reversible.
Fourth, although the case that the total energy of two colliding particles 
is zero is indeed extremely exotic, it is possible to customize the very 
problem 
setting with a tangible ground.
Let a particle  be capable of spontaneous decaying into two interacting
particles, with the total energy of this system being equal to 
zero. 
Then one  extends analytically this history back in time according to Eq.
(\ref{e-}).  
Of course, this trick only helps in rendering the ``real'' history of forming 
the aggregate of two particles a virtual history (which is to drop out of 
sight).
Hence, it may be argued that letting the existence of 
such point aggregates does not stand up.
The key step is to switch from particular aggregates to a continual set of 
identical aggregates constituting a string.
Leaving aside the origin of such sets, we 
take advantage of discrete toy models
of a string for better understanding the classical mechanism of its splitting. 

\section{Two-dimensional world}
We now consider classical electrodynamics in two-dimensional spacetime.
Our notations are identical to those of Ref. \cite{k2006}.
The action for a system of $N$ charged point particles and the 
electromagnetic field is given by
\begin{equation}
S=-\sum_{I=1}^N m_I\int ds_I\, \sqrt{{\dot z}_{\mu}^I\,{\dot z}^\mu_I}
-\int d^2 x\left(\frac18\,F_{\mu\nu}F^{\mu\nu}+ A^\mu j_\mu\right),
\label
{S}
\end{equation}                                           
\begin{equation}
j^\mu(x)=\sum_{I=1}^N e_I\int_{-\infty}^\infty ds_I\,{\dot z}^{\mu}_I(s_I)\,
\delta^{(2)}\left[x-z_I(s_I)\right].
\label
{j}
\end{equation}                                           
Here, the field strength $F_{\mu\nu}$ is related to the vector potential 
$A_\mu$ in the conventional way 
\begin{equation}
F_{\mu\nu}=\partial_\mu A_\nu-\partial_\nu A_\mu. 
\label
{F=dA}
\end{equation}                                           
Varying $A^\mu$ and $z^\mu_I$, we have
\begin{equation}
\partial_\lambda F^{\lambda\mu}=2 j^\mu,
\label
{maxw}
\end{equation}                                           
\begin{equation}
m_I{\ddot z}^{\mu}_I=e_I{\dot z}_{\alpha}^I F^{\mu\alpha}(z_I).
\label
{newt}
\end{equation}                                           

A  remarkable fact is that this system of equations is completely integrable
\cite{k2006}.
The procedure of finding solutions to this system is rather standard.
First we obtain a retarded solution to the field equation (\ref{maxw}) with
the source composed of $N$ charges moving along arbitrary smooth timelike
world lines.
The notation 
$R^\mu_I=x^{\mu}-z^{\mu}_{I}(s_I^{\rm ret})$
is used to denote
the 
null vector 
drawn
from the emission point $z^{\mu}_{I}(s_I^{\rm ret})$ on the $I$th world line 
to the point of observation $x^\mu$. 
From
here on the mark `ret' will be suppressed.
We introduce a further null vector $c^\mu_I$ related to $R^\mu_I$
by
\begin{equation}
R^\mu_I=\rho_I\,c^\mu_I,
\label
{def}
\end{equation}                                          
where 
\begin{equation}
\rho_I={\dot z}_I\cdot R_I 
\label
{rho-def}
\end{equation}                                          
is the
distance between emission and 
observation points in the frame in which the time axis is aligned 
with ${\dot z}^\mu_I$.
The retarded solution to (\ref{F=dA})--(\ref{maxw}) can be written \cite{k2006} as
\begin{equation}
A^{\mu}=-\sum_{I=1}^N e_IR_I^{\mu},
\label
{A}
\end{equation}                                             
\begin{equation}
F^{\mu\nu}=\sum_{I=1}^Ne_I\left(c_I^{\mu}{\dot z}_I^{\nu}
-c_I^{\nu}{\dot z}_I^{\mu}\right).
\label
{F}
\end{equation}                                             

As an illustration let us consider the case $N=2$.
This two-particle problem can be translated into the problem of motion of two
parallel plates of a planar immense capacitor.
Evidently there is only an electric field ${\bf E}$ between the plates, which is
constant for any separation and velocity of the plates.

Applying (\ref{F}) to the symmetric stress-energy tensor of the
electromagnetic field 
\begin{equation}
\Theta^{\mu\nu}=\frac12\left(F^{\mu\alpha}F_{\alpha}^{~\nu}+ 
\frac{\eta^{\mu\nu}}{4}\,F_{\alpha\beta}F^{\alpha\beta}\right)
\label
{Theta}
\end{equation}                                           
gives
$\Theta^{\mu\nu}_{\rm self}+
\Theta^{\mu\nu}_{\rm mix}$ where $\Theta^{\mu\nu}_{\rm self}$ is the 
sum of terms each containing only the field generated by a particular
charge, 
and $\Theta^{\mu\nu}_{\rm mix}$ contains mixed contributions.
Let $\Theta^{\mu\nu}_{I}$ be a term of $\Theta^{\mu\nu}_{\rm self}$ due to 
the $I$th charge.
We have  
\begin{equation}
\Theta^{\mu\nu}_{I}=
{\frac14}\,e_I^2\,\eta^{\mu\nu}.
\label
{Theta self}
\end{equation}                                           
This expression suggests that 
there is no radiation in two-dimensional spacetime (for an extended discussion 
of this subject see \cite{k9, k2006}.)

If we substitute (\ref{F}) in (\ref{newt}) and solve the resulting equations,
then we find that every particle moves along a hyperbolic
world line \cite{k2006}.
In ``degenerate'' cases, the history of a particle is represented by
straight world lines.

We now return to the system of two colliding particles, discussed in
the previous section.
For simplicity, we choose the barycentric frame, and assume that the particles 
have 
equal masses $m$ and charges $e$.
Our concern here is with the case that velocities of the particles are 
precisely zero at the instant of their meeting.
The exact solution  
\cite{Kosyakov2000} is represented by two world 
lines $z^\mu_1(s)$ and ${\ z}^\mu_2(s)$ which coalesce 
at $s=s^\ast$ and separate at $s=s^{\ast\ast}=s^{\ast}+T$, 
\begin{equation}
z^\mu_1(s)
=\cases
{a^{-1}\left({\sinh}\,a(s-s^\ast),1-{\cosh}\,a(s-s^\ast)\right)              
& $s<s^{\ast}$,\cr 
\left(s-s^{\ast},0\right)                                                
& $s^{\ast}\le s<s^{\ast\ast}$\cr 
a^{-1}\left(aT+{\sinh}\,a(s-s^{\ast\ast}),{\cosh}\,a(s-s^{\ast\ast})-1\right)
&$s\ge s^{\ast\ast}$\cr},  
\label
{z}
\end{equation}
\begin{equation}
z^\mu_2(s)
=\cases
{a^{-1}\left({\sinh}\,a(s-s^\ast),{\cosh}\,a(s-s^\ast)-1\right)
& $s<s^{\ast}$,\cr 
z^\mu_1(s)
&$s^{\ast}\le s<s^{\ast\ast}$,\cr 
a^{-1}\left(aT+{\sinh}\,a(s-s^{\ast\ast}),1-{\cosh}\,a(s-s^{\ast\ast})\right)
&$s\ge s^{\ast\ast}$.\cr}  
\label
{-z}
\end{equation}
Here, $a=e^2/m$.

The parameters $s^{\ast}$ and $s^{\ast\ast}$ are arbitrary. 
If $s^{\ast}$ and $s^{\ast\ast}$ are different and finite, then Eqs.
(\ref{z}) and (\ref{-z}) correspond to the history of an aggregate with
finite life time. 
If $s^{\ast\ast}\to\infty$, then this solution represents 
the history of a stable aggregate originated at a
finite instant.
In the limit $s^{\ast}\to -\infty$, we have 
the history of an aggregate, formed at the infinitely remote
past, whose decay occurs at a finite instant. 
If $s^{\ast}\to -\infty$ and $s^{\ast\ast}\to\infty$, then this solution
becomes a straight line
corresponding to an absolutely stable aggregate.
For $s^{\ast}=s^{\ast\ast}$, this solution describes an aggregate existing
for a single moment. 

We thus see that the exact solution
to the Cauchy problem for the set of equations 
governing a closed system of two charged particles and the electromagnetic 
field in two-dimensional spacetime, with 
the initial condition that the total energy of this system is 
zero, is not unique. 
In fact, we have a continuum of solutions (\ref{z})--(\ref{-z}) where $T$ is 
arbitrary: the aggregate disintegrates quite accidentally at
any instant after its formation\footnote{It may be worth noticing once again 
that 
the colliding particles cannot bounce off in the ordinary way 
because both velocities and interparticle repulsion vanish at 
the instant of their meeting $s=s^\ast$.}. 

Turning to fundamental strings, we begin to think of them as chain 
structures. 
An example is a system of two particles which are held together by the linearly rising 
potential (\ref{A}), resembling a string whose energy is linear in its 
length. 
While the particles exchange electromagnetic signals 
along 
the two-dimensional light cone, string perturbations (in the orthonormal gauge)
are 
governed by the wave equation 
$
{X}_{\tau\tau}-{X}_{\sigma\sigma}=0
$
whose characteristic surface is the light cone in the $(\tau,\sigma)$-plane.
If a two-parameter family of curves, labelled by 
$\tau$ and $\sigma$,  is drawn perpendicular to the world lines of the 
particles, then we have a toy discrete model of strings with Dirichlet 
boundary conditions 
$
{X}({\tau},0)={X}(\tau,l)=0$.

$N$-particle clusters with $N>2$ are also suitable for modeling such strings.
It is possible to follow the course of joining of two open strings into one and 
subsequent spontaneous splitting of this string into two pieces
if the extreme left particle of a cluster on the right and the 
extreme right particle of a cluster on the left move to meet 
(Fig. \ref{join-split}$a$) and merge into a single 
point aggregate  (Fig. \ref{join-split}$b$), and 
then, after a lapse
of a time interval $T$, this aggregate disintegrates into two initial 
particles (Fig. \ref{join-split}$c$), according to
Eqs.  (\ref{z})--(\ref{-z}).
One may then deem a classical string to be a set of aggregates of this 
kind.
Spontaneous disintegration of some element of this set is the reason for 
splitting of the string. 
\begin{center}
\begin{figure}[htb]
\unitlength=1.00mm
\special{em:linewidth 0.4pt}
\linethickness{0.4pt}
\begin{picture}(90.00,30.00)
\put(10.00,8.00){\makebox(0,0)[cc]{$\bullet$}}
\put(14.00,8.00){\makebox(0,0)[cc]{$\bullet$}}
\put(20.00,8.00){\makebox(0,0)[cc]{$\bullet$}}
\put(40.00,8.00){\makebox(0,0)[cc]{$\bullet$}}
\put(46.00,8.00){\makebox(0,0)[cc]{$\bullet$}}
\put(50.00,8.00){\makebox(0,0)[cc]{$\bullet$}}
\put(70.00,8.00){\makebox(0,0)[cc]{$\bullet$}}
\put(75.00,8.00){\makebox(0,0)[cc]{$\bullet$}}
\put(80.00,8.00){\makebox(0,0)[cc]{$\odot$}}
\put(80.00,8.00){\makebox(0,0)[cc]{$\bullet$}}
\put(85.00,8.00){\makebox(0,0)[cc]{$\bullet$}}
\put(90.00,8.00){\makebox(0,0)[cc]{$\bullet$}}
\put(110.00,8.00){\makebox(0,0)[cc]{$\bullet$}}
\put(116.00,8.00){\makebox(0,0)[cc]{$\bullet$}}
\put(120.00,8.00){\makebox(0,0)[cc]{$\bullet$}}
\put(140.00,8.00){\makebox(0,0)[cc]{$\bullet$}}
\put(144.00,8.00){\makebox(0,0)[cc]{$\bullet$}}
\put(150.00,8.00){\makebox(0,0)[cc]{$\bullet$}}
\put(20.00,13.00){\vector(1,0){4.00}}
\put(40.00,13.00){\vector(-1,0){4.00}}
\put(124.00,13.00){\vector(-1,0){4.00}}
\put(136.00,13.00){\vector(1,0){4.00}}
\put(30.00,2.00){\makebox(0,0)[cc]{$a$}}
\put(80.00,2.00){\makebox(0,0)[cc]{$b$}}
\put(130.00,2.00){\makebox(0,0)[cc]{$c$}}
\end{picture}
\caption{Joining and splitting of ``strings''}
\label
{join-split}
\end{figure}
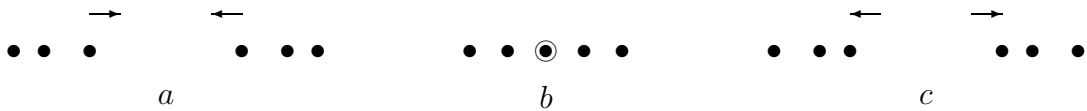 
\end{center}

One may wish to use the Yang--Mills theory with point
particles endowed with color charges transforming as the adjoint representation of 
the gauge group in constructing such string models as an alternative to 
electrodynamics.
But this analysis accomplishes nothing new:
all retarded solutions to the Yang--Mills equations in
two-dimensional spacetime are Abelian, that is, they can be built with the aid of
the Cartan subgroup of the gauge group \cite{k9}, so that we revert to the
situation in two-dimensional electrodynamics. 

\section{Refinement of the path-integral concept in classical 
mechanics}
The path-integral formulation 
of classical mechanics developed in \cite{gozzi}--\cite{persik} opened up 
new avenues 
for studies of the connection between the quantum and the classical.  
A central idea of this approach is that the classical path integral is 
contributed by 
a single path that renders the action extremal. 
Useful though this concept for truly 
deterministic systems, it must be
modified if we wish to incorporate systems violating Laplace's determinism.

It is common for the path-integral approach to take
the principle of least action in Hamilton's form
\begin{equation}
S=\int_{0}^T dt\,L(q,{\dot q}),
\quad
\hfill\hbox{or}
\quad
S=\int_{0}^T dt\left[p{\dot q}-H(q,p)\right].
\label
{Hamilton's-principle}
\end{equation}                                           
One may inquire: what is 
the  classical transition amplitude  
$K(\phi_f,T|\phi_i,0)$ of arriving at a phase space point 
$\phi_f=(q_f,p_f)$ at time $t_f=T$ having started from $\phi_i=(q_i,p_i)$ at
time $t_i=0$?  
The answer \cite{gozzi}--\cite{persik} is  
\begin{equation}
K(\phi_f, T|\phi_i,0)=\int [{{\cal D}}\phi]\,\,\delta(\phi-\phi_{\rm cl}).
\label
{gozzi-initial-PI}
\end{equation}                                           
Here, $[{{\cal D}}\phi]$ means integration is to be carried out in the 
space of all paths from $\phi_i$ to $\phi_f$, and $\phi_{\rm cl}$ is the 
extremal phase space path between $\phi_i$ to $\phi_f$.
Because
\begin{equation}
\delta(\phi-\phi_{\rm cl})
=
\delta\!\left(\frac{\delta S}{\delta \phi}\right)
\det\!\left[\frac{\delta^2 S}{\delta \phi(t')\delta \phi(t'')}\right],
\label
{delta=delta-det}
\end{equation}        
one may further take the Fourier transform of the Dirac delta and exponentiate
the determinant using  Grassmannian ghost variables $c$ and ${\bar c}$
to yield
\begin{equation}
K(\phi_f,T|\phi_i,0)=\int [{{\cal D}}\phi]\,
{\cal D}\lambda\,{\cal D}{\bar c}\,{\cal D}c\,
\exp\!\left(i\lambda \,\frac{\delta S}{\delta \phi}
+{\bar c}\,\frac{\delta^2 S}{\delta \phi^2}\,c\right).
\label
{gozzi-PI}
\end{equation}                                           
If we define two anticommuting partners of $t$, ${\bar\theta}$ and $\theta$, and
assemble the variables $\phi,\lambda,{\bar c},c$ into a single 
combination 
\begin{equation}
\Phi
=\phi
+{\bar\theta}{\bar c}
+\theta c
+i{\bar\theta}\theta\lambda,
\label
{supervariable}
\end{equation}                                           
then it is possible to rewrite (\ref{gozzi-PI}) in a very compact and 
elegant supersymmetric form \cite{persik}:
\begin{equation}
K(Q_f,T|Q_i,0)=\int [{{\cal D}}Q]\,{\cal D}\!P\,
\exp\!\left(-\int d{\bar\theta}\,d\theta\,\,S[\Phi]\right),
\label
{gozzi-PI-final}
\end{equation}                                           
which bears the formal similarity to the quantum path integral.

Recalling the particle moving to the top of a hill
on condition that the total energy is fixed to be zero, Eq. (\ref{hill}),
one is inclined to think of the principle of least action in Jacobi's form \cite{lanczos}
as a suitable starting point for
the description of indeterministic dynamics.  

We now outline general features of Jacobi's action\footnote{Our brief review is 
loosely patterned on the detailed exposition of Ref. \cite{brownjork}.}.
Consider a nonrelativistic system  described by a $n$-dimensional 
configuration space with
coordinates $q^a$, $a=1,2,\ldots,n$.
Let  this system be moving along a path $q^a(\sigma)$ whose argument 
$\sigma$ ranges  
from $0$ to $1$.
We denote ${q'}^a=dq^a/d\sigma$, and introduce the Newtonian metric $m_{ab}(q)$
(for a single point particle of mass $m$, with the use of Cartesian coordinates, $m_{ab}=
m\,\delta_{ab}$).
Jacobi's action is an integral over the configuration space trajectory,
\begin{equation}
{\bar S}=\int_{0}^{1} d\sigma\,\sqrt{m_{ab}(q)\,{q'}^a{q'}^b}\,\sqrt{
2\left[E-U(q)\right]},
\label
{Jacobi's-action}
\end{equation}                                           
where $U(q)$ is the potential energy.
In (\ref{Jacobi's-action}), the physical time interval between initial and 
final configurations is not fixed.
By contrast, the total energy of the system $E$ is fixed.

Varying (\ref{Jacobi's-action}) gives a trajectory  
$q(\sigma)$.
With the knowledge of $q(\sigma)$, it is possible to determine how the system 
evolves in time using a supplementary condition
\begin{equation}
\frac12\,m_{ab}\,{{\dot q}^a}\,{\dot q}^b+
U(q)=E,
\label
{suppl-cond}
\end{equation}                                           
where ${\dot q}^a=dq^a/dt$.

Since Jacobi's action (\ref{Jacobi's-action}) is invariant under the change of 
parametrization 
$\sigma\to f(\sigma)$ 
{with}
$f(0)=0$
{and}
$f(1)=1$,
the Hamiltonian associated with the Lagrangian in 
 (\ref{Jacobi's-action}) vanishes identically,
\begin{equation}
{\cal H}={q}^a\,\frac{\partial L}{\partial q^a}-L=0.
\label
{cal-H=0}
\end{equation}                                           
To put it differently, reparametrization invariance of the  action (\ref{Jacobi's-action})
leads to a constraint
\begin{equation}
{\cal H}(q,p)=\frac12\,m^{ab}\,p_a p_b+U(q)-E\approx 0,
\label
{cal-H-approx-0}
\end{equation}                                           
where $p_a$ is the momentum conjugate to the configuration coordinate $q^a$,
\begin{equation}
p_a
=
\frac{\partial L}{\partial {q'}^a}
=\frac{{q'}_a}{\sqrt{m_{ab}\,{q'}^a{q'}^b}}\,\sqrt{2\left[E-U(q)\right]}.
\label
{Jacobi's-momenta}
\end{equation}                                           

Because the canonical Hamiltonian ${\cal H}$ is zero, there are no secondary 
constraints, 
and ${\cal H}$ is trivially first class.
The action in canonical form is
\begin{equation}
{\bar S}
=\int_{0}^{1} 
d\sigma\left(p_{a}{q'}^a-N{\cal H}\right),
\label
{Jacobi's-action-canon}                                          
\end{equation}                                           
where $N$ is a Lagrange multiplier, whose variation enforces the constraint 
(\ref{cal-H-approx-0}).

The action (\ref{Jacobi's-action-canon}) is to be varied with $q(0)=q_i$
and $q(1)=q_f$ held fixed.
The equations of motion following from (\ref{Jacobi's-action-canon}) are 
\begin{equation}
{q'}^a=N p^a,
\quad
{p'}_a=-N\left(\frac12\,p_b p_c\,
\frac{\partial m^{bc}}{\partial {q}^a}
+ 
\frac{\partial U}{\partial {q}^a}\right),
\quad
\frac12\,p_a p^a+U(q)-E=0,
\label
{Jacobi's-eq-motion-canon}
\end{equation}                                           
where $p^a=m^{ab}p_b$.
Combining the first and third of these equations gives
\begin{equation}
N=
\left[\frac{{q'}^a{q'}_{\!a}}{2(E-U)}\right]^{\frac12}.
\label
{N=sqrt}
\end{equation}                                           
By (\ref{suppl-cond}),  
\begin{equation}
\frac{dt}{d\sigma}=
\left[\frac{{q'}^a{q'}_{\!a}}{2(E-U)}\right]^{\frac12}.
\label
{sqrt}
\end{equation}                                           
Therefore, $dt=Nd\sigma$, and so
\begin{equation}
T=\int^{1}_{0} d\sigma\,N(\sigma).
\label
{dt=Ndsigma}
\end{equation}                                           

This suggests that 
$N$ is the lapse in physical time associated with an increment in the
variable $\sigma$ parametrizing the phase space trajectory.
Note that this interpretation for $N$ will be maintained for as long as 
the flow of 
$t$ is correlated with increasing $\sigma$. 
This, however, 
is not the case for indeterministic regimes of evolution. 

The action (\ref{Jacobi's-action-canon}) is invariant under infinitesimal 
reparametrizations $\delta\sigma=\epsilon(\sigma)$ with
$\epsilon(0)=\epsilon(1)=0$
if one takes the transformation laws
\begin{equation}
\delta{q}=\epsilon q',
\quad
\delta{p}=\epsilon p',
\quad
\delta{N}=\left({\epsilon N}\right)',
\label
{reparam-transf}
\end{equation}                                           
which are generated by the first class constraint (\ref{cal-H-approx-0}).

If we express the momenta $p$ in terms of the velocities ${q'}$,  then
(\ref{Jacobi's-action-canon}) becomes
\begin{equation}
{\bar S}
=\int_{0}^{1} d\sigma\left[\frac{m_{ab}\,{q'}^a{q'}^b}{2N}+
N\left(E-U\right)\right].             
\label
{Jacobi's-action-lagr}
\end{equation}                                           
Integrating away $N$ from (\ref{Jacobi's-action-lagr}), we return to
the original Jacobi action (\ref{Jacobi's-action}).

We now focus on 
one-dimensional systems.
We first define an invariant path-integral measure  for deterministic 
reparametrization-invariant systems\footnote{Subsequent to posting the 
first version of this paper, I was told by Prof. E. Gozzi that the classical 
path integral with reparametrization invariance was handled by the 
Batalin--Fradkin--Vilkovisky
method in Ref. \cite{thacker}.
The result of this work provides an alternative, but equivalent to (\ref{pth-int-mod}),  
formulation of  the classical 
path integral using Jacobi's action.}.
Following \cite{polyakov,mottola}, we integrate over the coset space of 
all functions $\phi(\sigma)$  and one-dimensional 
metrics $N(\sigma)$ modulo reparametrizations, 
\begin{equation}
\frac{{\cal D}\phi(\sigma)\,{\cal D} N(\sigma)}{{\cal D} f(\sigma)}\,,
\label
{path-int-measure}
\end{equation}                                           
where
${\cal D} N(\sigma)/{{\cal D}f(\sigma)}$  
can be shown  \cite{polyakov,mottola} to
reduce to a
conventional Lebesgue measure $dT$, with 
$T$ being the physical time interval given by (\ref{N=sqrt}).
By applying these results to the procedure of Ref. \cite{persik}, we 
recast 
(\ref{gozzi-PI-final}) 
in the form
\begin{equation}
K(Q_f|Q_i)
=\int_0^\infty dT\int [{{\cal D}}Q]\, {\cal D}\!P\,
\exp\left[-\int_{0}^{1} d\sigma\, d{\bar\theta}\,d\theta
\left(P{Q'}-T{\cal H}\right)\right],
\label
{pth-int-mod}
\end{equation}                                           
where the conjugate  
supervariables  $Q$ and $P$ are patterned after Eq. (\ref{supervariable}),
and
\begin{equation}
{\cal H}(Q,P)=\frac{1}{2m}\,P^{2}+U(Q)-E.
\label
{cal-H(Q,P)}
\end{equation}

A modifications of the path integral 
for indeterministic systems can be ascertained by the example of a particle 
that moves to the top of the hill $U(q)=-U_0\,|q|$, equilibrates at $q=0$ for 
an arbitrary period of time $T$, and then descends down the hill.
The initial and final stages of this
process, that is, the ascent and descent, are essentially deterministic.
Hence, the  classical transition amplitude  
for these stages is deduced from
(\ref{pth-int-mod})--(\ref{cal-H(Q,P)}).

Care must be exercised in treating the indeterministic stage---that is, 
the stay at the top.
Let us assume that $T$ is a discrete variable taking values $0,\ell,2\ell,\ldots$
The {prior} probability that the particle will be at rest after 
completing one quantum of time $\ell$ is $\frac12$.
After a lapse of two quanta of time $2\ell$, this quantity 
is $\left(\frac12\right)^2$.
And so on. 
With this assumption, the $T$-integration is substituted for a discrete sum,
and hence
\begin{equation}
K(\phi_f=0|\phi_i=0)
=\left(\frac12
+\frac{1}{4}+\ldots\right)\int [{\cal D} q]\,[{\cal D} p] \,
\delta(q)\, 
\delta(p)
=\frac{1}{16}.
\label
{left-int-cal-D-q-D-p}
\end{equation}                                           
Here, the end-point integrals of the Dirac deltas over half-infinite intervals
are understood as appropriate limits of integrals of
sequences of functions, such as
\begin{equation}
\int_0^\infty dx\,\delta(x)\varphi(x)
=\lim_{\epsilon\to 0}\,\frac12\int_{-\infty}^{\infty}dx\,
\frac{{\epsilon}}{{\pi}\left(x^2+\epsilon^2\right)}\,\varphi(x)
=\frac12\,\varphi(0).
\label
{int-delta}
\end{equation}

It would be interesting to see if 
it is possible to bridge this random
dynamics 
arising from
spontaneous equilibrium breaking 
with that owing its origin to  
't~Hooft's information loss 
condition (for a discussion of this condition  and further references see
\cite{blasone}).

\section*{Acknowledgments}
I am indebted to Prof. John Norton for pointing out Ref. \cite{norton} and 
interesting comments.
I am especially thankful to Prof. Ennio Gozzi for many valuable discussions.

\end{document}